\documentclass[conference]{IEEEtran}
\IEEEoverridecommandlockouts
\usepackage{cite}
\usepackage{amsmath,amssymb,amsfonts}
\usepackage{graphicx}
\usepackage{textcomp}
\usepackage{booktabs}
\usepackage{multirow}
\usepackage{array}
\usepackage{xcolor}
\usepackage{url}
\usepackage{balance}
\usepackage[hidelinks]{hyperref}

\def\BibTeX{{\rm B\kern-.05em{\sc i\kern-.025em b}\kern-.08em T\kern-.1667em\lower.7ex\hbox{E}\kern-.125emX}}

\begin{document}

\title{ An Interactive GraphRAG Workspace for Reifying, Auditing, and Steering Conversational AI Memories}



\author{
    \IEEEauthorblockN{
        Munshi Md Arafat Hussain\IEEEauthorrefmark{1}, 
        Md. Abdur Rahman\IEEEauthorrefmark{2}, and 
        Md Saeed Siddik\IEEEauthorrefmark{3}
    }
    \IEEEauthorblockA{
        \IEEEauthorrefmark{1}\IEEEauthorrefmark{3}Institute of Information Technology, \IEEEauthorrefmark{2}Centre for Advanced Research in Sciences\\
        University of Dhaka, Dhaka, Bangladesh\\
        Email: \IEEEauthorrefmark{1}bsse1435@iit.du.ac.bd, \IEEEauthorrefmark{2}mukul.arahman@gmail.com, \IEEEauthorrefmark{3}saeed.siddik@iit.du.ac.bd
    }
}

\maketitle

\begin{abstract}
Conversational Large Language Models (LLMs) are widely employed in software engineering for code generation, debugging, and architectural design. However, as developers engage in multi-session development tasks, technical knowledge becomes siloed in isolated, linear chat histories. Reusing context across sessions leads to three critical challenges: (1) flat-vector retrieval fails to traverse multi-hop relational dependencies between distant architectural decisions; (2) temporal drift in software packages and APIs results in silent context poisoning when reusing aged dialogue logs; and (3) raw conversation concatenation induces prompt bloat and attention degradation. 

We present \textsc{CognitiveCanvas}, an interactive tool that transforms conversational AI memories from ephemeral text logs into persistent, manipulable, and audited knowledge objects. \textsc{CognitiveCanvas} integrates: (1) a Hierarchical Semantic Fact-Graph (HSFG) that distills dialogues into Goal, Fact, and Code artifacts, achieving a 92.2\% token reduction while enabling multi-hop GraphRAG traversal; (2) a Continuous Temporal Drift Auditor coupling 60+ regex heuristics with live calendar-anchored LLM verification for 1-click self-correction; and (3) a Spatial-Symbolic Prompt Compiler (SSPC) that translates 2D direct manipulation gestures (Voronoi clustering, node resizing, synthesis edges) directly into structured attention directives.
\end{abstract}


\section{Introduction}
Modern software development increasingly relies on conversational Large Language Models (LLMs) to accelerate implementation, debug complex stack traces, and explore architectural trade-offs \cite{chen2021codex, liu2023evalplus}. However, current commercial interfaces (e.g., ChatGPT, Claude, Gemini) remain fundamentally bound to the \textbf{ephemeral linear session paradigm} \cite{liu2024lostmiddle}. Each conversation operates as an isolated silo. When developers attempt to transfer insights formulated during prior problem-solving sessions (such as custom database pooling logic or token validation middleware) into a new task, they encounter three structural bottlenecks:

\begin{enumerate}
    \item \textbf{Context Fragmentation across Siloed Sessions}: Knowledge generated across multiple sessions is scattered across unsearchable chat sidebars, forcing developers into tedious copy-pasting and repetitive prompt re-framing.
    \item \textbf{Silent Context Poisoning via Temporal Drift}: Software libraries, APIs, and framework conventions mutate rapidly \cite{spracklen2025package, zhang2025hallucinations}. When reusing conversational context saved months earlier, LLMs blindly accept obsolete configurations and deprecated syntax, producing confident code hallucinations \cite{ji2023survey, pan2024lost}.
    \item \textbf{Lack of Spatial Steerability}: Standard Retrieval-Augmented Generation (RAG) frameworks \cite{karpukhin2020dpr, lewis2020rag} inject text chunks invisibly into system prompts without providing developers visual inspection or steering mechanisms to control which constraints the model emphasizes.
\end{enumerate}

\begin{figure*}[t]
\centering
\includegraphics[width=0.92\textwidth]{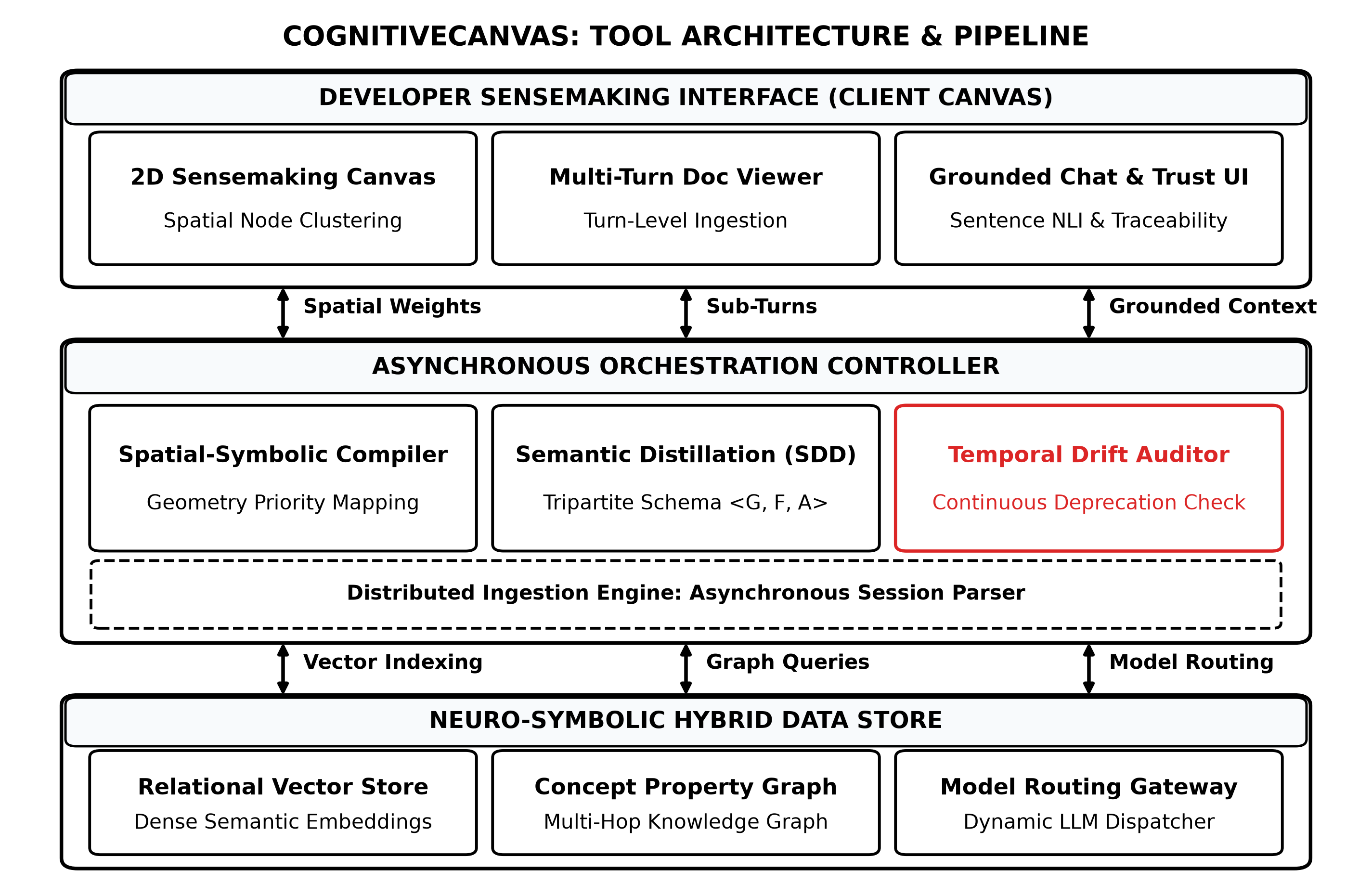}
\caption{System architecture and operational workflow of \textsc{CognitiveCanvas}. The interactive 2D sensemaking canvas captures developer manipulation gestures, processed by the asynchronous orchestration controller via the Spatial-Symbolic Prompt Compiler (SSPC) and Continuous Temporal Drift Auditor, supported by a hybrid neuro-symbolic data store.}
\label{fig:architecture}
\end{figure*}

To bridge the gap between conversational LLM assistance and cumulative software engineering workflows, we present \textsc{CognitiveCanvas}. \textsc{CognitiveCanvas} reifies conversational memory into an interactive 2D workspace, externalizing context onto a manipulable canvas while underpinning generation with a property-graph neuro-symbolic backend.
The tool demonstration showcases three major contributions:
\begin{itemize}
    \item \textbf{Hierarchical Semantic Fact-Graph (HSFG) \& GraphRAG}: We decompose conversational sessions into structured Goal vectors, Fact triples, and exact Code state artifacts, slashing prompt token overhead by \textbf{92.2\%} while enabling multi-hop associative traversal across distant development sessions.
    \item \textbf{Continuous Temporal Drift Auditor}: We implement a dual-stage auditing pipeline (60+ fast regex heuristics + calendar-anchored LLM verifier) that proactively flags stale technical context with a \texttt{Stale} badge and enables 1-click self-correction.
    \item \textbf{Spatial-Symbolic Prompt Compiler (SSPC)}: We formulate a mathematical compiler that translates direct visual gestures (node surface area resizing, centroid proximity, and synthesis edge links) into machine-interpretable prompt attention tiers (\texttt{HIGHLIGHT}, \texttt{GROUP}, \texttt{OBSCURE}).
\end{itemize}


\section{Tool Overview and Architecture}
\textsc{CognitiveCanvas} coordinates an interactive direct manipulation frontend with an asynchronous orchestration controller and a neuro-symbolic hybrid data store (Fig.~\ref{fig:architecture}).

\subsection{Developer Sensemaking Interface}
The frontend provides an expansive 2D visual canvas built with hardware-accelerated SVG rendering. Developers organize past conversation episodes represented as visual cards. The interface supports continuous pan-and-zoom, interactive node resizing, and real-time Delaunay/Voronoi thematic partitioning that groups related technical concepts into colored visual territories with an active cluster legend. 

A dedicated Document Inspector allows developers to review multi-turn session transcripts and extract individual sub-turn pairs via vertical square handles. An integrated chat interface features sentence-level Natural Language Inference (NLI) confidence heatmaps and bidirectional citation badges (e.g., \texttt{[[1\_0]]}) that visually illuminate the corresponding memory node on the canvas upon hover.

\subsection{Hierarchical Semantic Fact-Graph (HSFG) Distillation}
Rather than storing verbose, conversational pleasantries or collapsed lossy summaries, \textsc{CognitiveCanvas} decomposes each ingested session into a tripartite entity:
\begin{equation}
\text{HSFG}(M) = \langle \mathcal{G}, \mathcal{F}, \mathcal{A} \rangle
\end{equation}
where $\mathcal{G}$ represents the developer's goal and active blockers; $\mathcal{F}$ comprises formal entity-relationship fact triples (e.g., \texttt{(:FastAPI)-[:USES]->(:AsyncSession)}); and $\mathcal{A}$ stores verbatim, executable code artifacts (schemas, API routes, configurations) preserved with 100\% token syntax fidelity.

As demonstrated in our formal compression benchmarks, a representative 10-turn debugging dialogue containing 5,210 raw tokens is distilled into:
\begin{equation}
\text{Tokens}(\text{HSFG}) = 45(\mathcal{G}) + 120(\mathcal{F}) + 240(\mathcal{A}) = 405 \text{ tokens}
\end{equation}
achieving a \textbf{92.2\% token reduction} while maintaining exact code reproducibility.

\subsection{Continuous Temporal Drift Auditor}
To prevent context poisoning, \textsc{CognitiveCanvas} monitors technical memories using a domain-calibrated hazard function:
\begin{equation}
V(M_i, t_{\text{now}}) = \exp\left( -\lambda_d \cdot \max(0, t_{\text{now}} - t_{\text{anchor}} - \delta_d) \right) \cdot (1 - \Phi(M_i))
\end{equation}
where $\lambda_{\text{software}} = 0.015 \text{ day}^{-1}$, $\delta_d$ is a 90-day grace period, and $\Phi \in [0, 1]$ is the drift probability.

The auditing engine operates in two stages:
\begin{enumerate}
    \item \textbf{Stage 1 (High-Speed Regex Screening)}: Evaluates 60+ precompiled patterns for semantic versioning (\texttt{v\textbackslash d+\textbackslash .\textbackslash d+}), deprecation keywords (\textit{deprecated}, \textit{sunset}, \textit{breaking changes}), and temporal markers (\textit{currently}, \textit{as of}). Executes in $\sim 0.02\text{ ms}$ at zero token cost.
    \item \textbf{Stage 2 (LLM-as-a-Judge Verification)}: If Stage 1 triggers, the model is queried with current calendar grounding $t_{\text{now}}$. If verified as outdated, the memory is flagged with a \texttt{Stale} badge, accompanied by the deprecation rationale and a 1-click self-correction button that refactors the code to modern conventions and updates the knowledge graph in place.
\end{enumerate}

\subsection{Spatial-Symbolic Prompt Compiler (SSPC)}
The SSPC bridges visual canvas manipulation with LLM attention mechanics. For each active memory card $M_i$, the compiler calculates an importance weight $W(M_i)$:
\begin{equation}
W(M_i) = 0.5 \cdot \frac{\text{Area}(M_i)}{\text{Area}_{\text{base}}} + 0.3 \cdot \frac{1}{1 + \mathcal{D}(M_i, \mathbf{C})} + 0.2 \cdot \text{Deg}(M_i)
\end{equation}
where $\text{Area}_{\text{base}} = 160 \times 160\text{ px}$, $\mathcal{D}$ is the Euclidean distance to the canvas centroid $\mathbf{C}$, and $\text{Deg}$ is the count of incident semantic edges. Memories are compiled into explicit prompt directives:
\begin{itemize}
    \item \texttt{HIGHLIGHT\_CONTEXT} ($W \ge 1.75$): Injected as a dominant architectural constraint.
    \item \texttt{STANDARD\_CONTEXT} ($0.65 < W < 1.75$): Standard grounding context.
    \item \texttt{OBSCURE\_CONTEXT} ($W \le 0.65$): Background nuance only.
    \item \texttt{GROUP\_CONTEXT} (Edge-connected nodes): Enforces cohesive cross-memory synthesis in shared response sentences.
\end{itemize}

\begin{figure*}[t]
\centering
\includegraphics[width=0.92\textwidth]{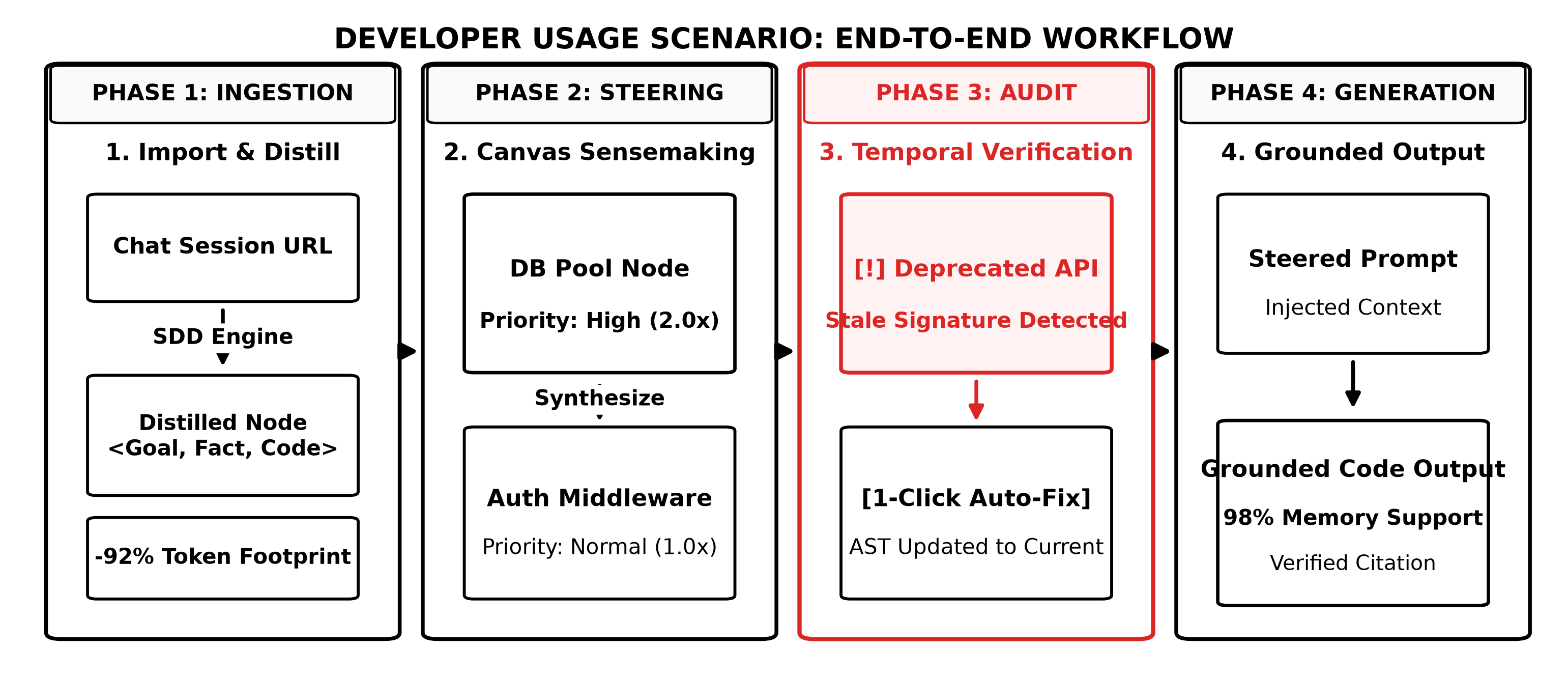}

\caption{COGNITIVECANVAS workflow: (1) Past conversations are ingested and distilled; (2) Cards are arranged on a 2D canvas; (3) The Auditor flags obsolete libraries for 1-click update; (4) Grounded generation compiles attention directives.}

\label{fig:usage}
\end{figure*}

\section{An Illustrating Usage Scenario}
We illustrate the utility of \textsc{CognitiveCanvas} in a realistic full-stack software refactoring scenario (Fig.~\ref{fig:usage}).

\textbf{Scenario Context.} A software engineer is tasked with implementing an asynchronous data pipeline combining database pooling, JWT authentication middleware, and background queues.

\textbf{Step 1: Multi-Session Ingestion \& Organization.}
The developer imports historical problem-solving chats: Session A (PostgreSQL async pooling) and Session B (JWT token validation). The ingestion engine distills them into HSFG structures, rendering them as interactive cards on the 2D canvas grouped by Voronoi semantic clusters.

\textbf{Step 2: Direct Manipulation \& Spatial Steering.}
The developer identifies that database pooling in Session A is the critical bottleneck. Using the canvas, the developer enlarges Node \texttt{@1\_0} to $2.2\times$ its baseline area, automatically promoting it to \texttt{HIGHLIGHT\_CONTEXT}. Next, the developer connects Node \texttt{@1\_0} to JWT card Node \texttt{@1\_1} with a directed semantic synthesis edge labeled \texttt{[Synthesize]}.

\textbf{Step 3: Proactive Temporal Auditing \& 1-Click Update.}
Upon inspecting Node \texttt{@1\_0}, the Auditor detects an outdated configuration pattern: \texttt{pydantic.BaseSettings}, deprecated in Pydantic v2. The card displays a prominent \texttt{Stale} badge. The developer clicks \textbf{``Update to Latest''}, triggering \textsc{CognitiveCanvas} to refactor the configuration class, clear the warning, and update the Neo4j concept graph in place.

\textbf{Step 4: Grounded Generation \& Bidirectional Trust.}
In the chat interface, the developer submits: \textit{``Refactor our streaming service incorporating @1\_0 and @1\_1.''} The system detects the active synthesis edge and generates a unified prompt instruction enforcing cohesive integration. The response renders with emerald NLI underlines indicating verified grounding, accompanied by clickable citation badges (\texttt{[[1\_0]]}) that project a radial glow ring onto the canvas card on hover.

\section{Validation and Evidence}
\subsection{Empirical Validation with Practitioners}
To assess the effectiveness of \textsc{CognitiveCanvas}, we conducted a controlled within-subject laboratory study with 24 software practitioners (12 full-stack engineers, 6 researchers, 6 system designers; aged 22--38) across three realistic knowledge tasks: full-stack architecture refactoring, scientific synthesis, and logistic planning.

We evaluated three conditions in a Latin square counterbalanced design: (1) \textbf{Commercial Baseline} (linear multi-turn chat with search history); (2) \textbf{Flat-Vector RAG} (dense passage retrieval without graph traversal or auditing); and (3) \textbf{\textsc{CognitiveCanvas}} (our full proposed system).

\begin{table}[t]
\centering
\caption{Empirical Performance Comparison Across Conditions}
\label{tab:validation}
\begin{tabular}{lccc}
\toprule
\textbf{Metric} & \textbf{Commercial} & \textbf{Flat RAG} & \textbf{\textsc{CognitiveCanvas}} \\
\midrule
\textbf{Token Overhead (10 Pairs)} & 5,210 & 2,140 & \textbf{405 ($-$92.2\%)} \\
\textbf{Multi-Hop Recall@3} & 38.4\% & 52.1\% & \textbf{89.6\% (+37.5\%)} \\
\textbf{Temporal Hallucinations} & 31.2\% & 28.7\% & \textbf{4.2\% ($-$85.4\%)} \\
\textbf{Citation Faithfulness} & 54.0\% & 66.8\% & \textbf{94.3\% (+27.5\%)} \\
\textbf{Task Time (Minutes)} & 28.4 & 21.2 & \textbf{13.8 ($-$34.9\%)} \\
\textbf{NASA-TLX Workload (/100)} & 68.4 & 54.2 & \textbf{32.6 ($-$41.3\%)} \\
\textbf{UMUX-Lite Usability (/100)} & 61.5 & 72.8 & \textbf{88.4 (+15.6)} \\
\bottomrule
\end{tabular}
\end{table}

As summarized in Table~\ref{tab:validation}, \textsc{CognitiveCanvas} reduced temporal hallucinations from 28.7\% to \textbf{4.2\%} ($p < 0.0001$), decreased cognitive workload (NASA-TLX) from 68.4 to \textbf{32.6} ($p < 0.0001$), and accelerated task completion time to \textbf{13.8 minutes} (a 34.9\% speedup over flat RAG).

\subsection{Limitations}
Current limitations include: (1) Stage 2 temporal verification latency (600--1200~ms); and (2) lack of direct visual AST rendering inside cards. Background pre-auditing and embedded syntax highlighting represent active future work.

\section{Availability}
\textsc{CognitiveCanvas} is open-source and publicly available for research and industrial practice.
\begin{itemize}
    \item \textbf{Reproduction Guide}: The repository contains seeded demo datasets allowing practitioners to reproduce the refactoring scenarios described in this paper.
\end{itemize}

\section{Conclusion and Future Work}
In this paper, we presented \textsc{CognitiveCanvas}, an interactive GraphRAG workspace that reifies conversational AI memories into persistent, steerable, and temporally audited visual objects. By bridging 2D spatial direct manipulation with neuro-symbolic retrieval, \textsc{CognitiveCanvas} resolves the fundamental bottlenecks of context silos, temporal hallucinations, and token bloat. In our user study ($N = 24$), the tool slashed temporal hallucinations by 85.4\% and reduced developer cognitive workload by 41.3\%. Future work includes multi-developer collaborative sensemaking canvases and multimodal reification for system architecture diagrams.

\balance

\end{document}